\documentclass[aps,preprint,amsfonts,amssymb,amsbsy,floats,graphicx,epsfig,tabularx]{revtex4}

\usepackage[dvips]{graphicx}
\newcommand{\be}{\begin{equation}}
\newcommand{\ee}{\end{equation}}
\newcommand{\p}{\partial}
\newcommand{\dd}{\delta}
\newcommand{\sa}{\sigma}
\newcommand{\prll}{Phys. Rev. Lett. }
\newcommand{\prbb}{Phys. Rev. B }

\newcommand{\jcpp}{J. Chem. Phys.}

\begin{document}
\title{A many-body interatomic potential for ionic systems: application to MgO\footnote{To appear in the Journal of Chemical Physics, October 2003}}
\author{P. Tangney\footnote{Author to whom correspondence should be addressed. Current address :
Dept. of Physics, University of California at Berkeley, Berkeley, CA 94720. Electronic mail : tangney@civet.berkeley.edu} and S. Scandolo}

\address{
International School for Advanced Studies (SISSA/ISAS), 
         via Beirut 2-4, 34013 Trieste, Italy.\\
INFM/Democritos, National Simulation Center, via Beirut 2-4, 34013 Trieste, Italy}
\date{\today}
\begin{abstract}
An analytic representation of the short-range repulsion energy in ionic systems is described that
allows for the fact that ions may change their size and shape depending on their environment.
This function is extremely efficient to evaluate relative to previous methods of modeling 
the same physical effects.
Using a well-defined parametrization procedure we have obtained parameter sets for this energy
function that reproduce closely the density functional theory potential energy surface of bulk
MgO. We show how excellent agreement can be obtained with experimental measurements of phonon
frequencies and temperature and pressure dependences of the density by using this effective potential
in conjunction with {\em ab initio} parametrization.
\end{abstract}
\maketitle
\section{Introduction}
The problem of modelling the dynamics of ionic materials has a long history\cite{sangdix,stoneham_and_harding}.
For many years the only available models were empirical force fields that were not very accurate 
or transferable between different environments. This was a problem both of the form of the model potentials
used and the way in which these potentials were parametrized. Most frequently, simple pairwise effective potentials
were used whose parameters were obtained using a combination of physical reasoning and empiricism. Such potentials
could, at best, only be expected to produce qualitative or poorly-quantitative results.

The advent of {\em ab initio} molecular dynamics\cite{carpar,paynermp} (MD) brought about a dramatic improvement
in the accuracy with which the potential energy surface of the ions could be calculated,
but this came with the price of an enormous increase in computational expense.
In molecular dynamics simulations the precision with which thermodynamic properties can be calculated 
depends on the size of the system studied and the length of the simulation over which averages may be taken.
For {\em ab initio} molecular dynamics one is generally confined to systems of around $100$ atoms and 
simulation times of $\sim 10$ picoseconds and so the precision with which many properties may be calculated
is poor. In addition, for highly viscous liquids (such as silica\cite{trave,trave2}) or very harmonic crystals the timescales
available within {\em ab initio} MD may not be sufficient to adequately equilibrate the system\cite{us_silica}. 
Some {\em ab initio}-sized systems may suffer from finite size effects.
For all these reasons there are many applications (good examples being
the determination of melting temperatures or finite-temperature elastic constants) that are very difficult to tackle with {\em ab initio} MD
and others (such as thermal conductivity and viscosity) that cannot be addressed at all.

Effective potentials are far less computationally expensive and so allow much longer simulation times and larger
system sizes. It is therefore extremely desirable to find a compromise between the accuracy of {\em ab initio} MD
and the computational speed of MD using effective potentials.

It has been shown\cite{furio,laio,madden_mgo2,us_silica} how the accuracy of an effective potential can 
be greatly improved by parametrizing using information from {\em ab initio} calculations.  However 
in order for this to work a physically appropriate form
for the potential must be used. Improving effective potentials therefore involves both using good parametrization
procedures and appropriate forms for the potentials. The form that an effective potential takes should be flexible enough
to describe the potential energy surface and specific enough to allow for efficient parametrization and application.

In this paper we present a many-body force field for ionic systems that incorporates the effect of an ion's
environment on its shape and size and the impact that such ionic distortions have on the short-range repulsive interactions between
ions. These effects have been shown in the past to be important to the bonding of many simple ionic 
systems\cite{schroder,madden_compress,madden_mgo1}.
The force field presented has much in common with previously proposed ``compressible-ion''\cite{madden_compress}
and ``aspherical-ion''\cite{madden_mgo1}
models but is superior from a computational point of view and avoids some problems associated with the way
in which these models are implemented (see section~\ref{section:manybody}).
It can also be implemented within a general 
mathematical framework that is very amenable to changes for the purposes of experimentation and improvement.
A particularly attractive feature of the proposed force-field is that it can describe distortions of arbitrary
shape. Previous models have been confined to describing distortions of certain given symmetries.

We apply the proposed model to crystalline and liquid magnesium oxide.
MgO has been used in the past as a testing ground for models of ionic systems\cite{madden_compress,madden_mgo1}.
It is considered to be the simplest oxide and yet it is a system in which many-body interactions
associated with distortions of the large oxide anion are known to be important\cite{schroder,boyer,madden_compress}. 
It is therefore a starting point for attempts to model the many body interactions in oxides and ionic systems
in general. 

MgO is a system of geophysical importance. It is an important component of the earth's lower mantle and
its stability up to high pressures\cite{duffy} make it useful as a pressure calibration standard for high pressure
and temperature experiments. Since the vast majority of the compounds present in the earth's mantle are oxides, MgO
is a starting point for studies of the effects of pressure and temperature on mantle materials.

For MgO the use of our model is found to result in a significant improvement (over pairwise representations of the 
short-range repulsion) in the ability to fit the forces, stresses, and energies 
extracted from density functional theory (DFT)\cite{dft} calculations.
We use the model in conjunction with electrostatic interactions (involving both charges and induced dipoles)
within a well-defined parametrization procedure. The resulting force field is shown
to give a very good description of the structure and dynamics of MgO.

\section{Parametrizing a force field using {\em ab initio} data}\label{section:param}
As mentioned above, it has been shown for a number of systems 
that a very high level of accuracy may be achieved by parametrizing an effective potential by fitting to 
DFT forces, stresses, and energies in selected atomic configurations\cite{furio,laio,madden_mgo2,us_silica}. 
However empiricism is also necessary unless one 
uses a functional form that is physically appropriate in the sense that it describes the electronic effects that are most important
for the property one is studying\cite{us_silica}. In particular in reference \onlinecite{us_silica} we have shown that, for silica, 
the inadequacy of simple pair potentials means that improving agreement
with the forces and stresses from {\em ab initio} simulations, in general, {\em disimproves} the ability of empirically derived 
parameter sets to describe crystal structures. By inclusion of more many-body effects (in this
case the polarizability of the oxygen anion) the ability of the force-field to reproduce {\em ab initio} 
data is improved and the empiricism becomes unnecessary for describing structure very accurately.

Here we make the assumption that {\em ab initio} calculations are highly accurate and always superior
to calculations using effective potentials.
This assumption is clearly formally unjustifiable but it is
useful for our purposes. We aspire to {\em ab initio}
accuracy and leave the improvement of the {\em ab initio} calculations
themselves as a separate problem.
We assume that if one achieves a perfect fit to {\em ab initio}
data one gets an extremely accurate effective potential. However, as one moves away from this limit, the relationship
between the fit to the {\em ab initio} data and the quality of the effective potential clearly weakens. If the fit is very
poor then (as with the description of silica using pair potentials\cite{us_silica}) 
even large improvements to it may not improve the ability of the potential 
to describe physical properties.

We define the fit as ${\bf \gamma}=(\Delta F,\Delta S,\Delta E)$, where $\Delta F$,$\Delta S$ 
and $\Delta E$ are dimensionless quantities corresponding to the average\cite{sample}
percentage differences between forces, stress components, and energy
differences between different configurations calculated {\em ab initio} and with the
effective potential, i.e.  
\begin{eqnarray}
\Delta F & = & 100\times\frac{\sqrt{\sum_{k=1}^{n_{c}}\sum_{I=1}^{N}\sum_{\alpha} |F_{ep,I}^{\alpha}(\{\eta\})
-F_{ai,I}^{\alpha}|^{2}}}
{\sqrt{\sum_{k=1}^{n_{c}}
\sum_{I=1}^{N}\sum_{\alpha} (F_{ai,I}^{\alpha})^{2}}}  \nonumber \\
\Delta S & = &
100\times\frac{\sqrt{\sum_{k=1}^{n_{c}}\sum_{\alpha,\beta} |S_{ep}^{\alpha\beta}(\{\eta\})-S_{ai}^{\alpha\beta}|^{2}}}
{3B\sqrt{n_{c}}} \nonumber \\
\Delta E & = &
100\times\frac{ \sqrt{\sum_{k,l}^{n_{c}} ((U^{k}_{ep}-U^{l}_{ep})-(U^{k}_{ai}-U^{l}_{ai}))^{2}}}
{\sqrt{\sum_{k,l}^{n_{c}}(U^{k}_{ai}-U^{l}_{ai})^{2}}} \nonumber
\end{eqnarray}
where $F^{\alpha}_{I}$ is the $\alpha^{\text{th}}$ cartesian component of the forces on ion $I$, 
$S^{\alpha\beta}$ denote stress components, $U^{k}$ is the potential energy of configuration $k$, 
$\{\eta\}$ is the set of parameters that, along with the functional form, characterise the 
effective potential. $B$ is a pressure that may be taken to be the bulk modulus of the material.
The superscripts 'ep' and 'ai' indicate whether a quantity has been calculated with the effective potential
or {\em ab initio}. In order for ${\bf \gamma}$ to be meaningful it is necessary that it be converged with 
respect to the number of configurations $n_{c}$ that have been used to test the effective potential. 
Typical values of $n_{c}$ are of the order of $10$. 

Since our parametrization scheme does not discriminate between different contributions
to the DFT forces, in order to be sure {\em a priori} that a potential
works well one must have a very high quality fit to the {\em ab initio} data.
This is because different properties depend to varying degrees on different contributions to the forces on the ions.
It is difficult to know how a given value of, for example, $\Delta F$, 
manifests itself in the property that one is interested in studying.
This is particularly true when there are a number of important contributions to the forces 
as these different contributions
may be fit by an effective potential to varying degrees.
If one has a potential that fits the {\em ab initio} forces perfectly then, clearly, each contribution is fit perfectly and
this uncertainty is eliminated.
For example, for silica the structure was extremely well described by using a good description of electrostatic 
effects\cite{us_silica}. However the forces differed
from the {\em ab initio} ones by $\sim 16 \%$. This suggests that the gross features of the potential energy surface are
well described but that more local features and finer details, that are important for dynamics, may not be. 
We have observed that the diffusion of liquid silica at $3500$ K as described by the model proposed in reference ~\onlinecite{us_silica}
seems to be too fast compared to extrapolations of lower temperature experimental data\cite{diffusion}. We also observe 
the phase transition between $\alpha-$ and $\beta-$cristobalite at too low a temperature. Both of these observations
are consistent with an underestimation of {\em local} energy barriers due to a functional form that 
oversimplifies the short-ranged interactions between ions.

For every given functional form there are minimum values of $\Delta F$,$\Delta S$, and $\Delta E$ in parameter space. 
Functional forms that oversimplify
the description of the electronic effects that are relevant for a given physical property
may not be capable of achieving a close enough fit to the {\em ab initio}
data to ensure that improvement of this fit improves the ability of the parameter set to describe physical properties.
If this is the case then it is still conceivable that one may improve the agreement with experiment on many physical
properties in an empirical way. We are of the opinion that an inability of a functional form 
to fit {\em ab initio} data indicates an intrinsic inadequacy of this form and that potentials created in this way should not 
be relied upon as being predictive. 
Effective potentials represent electronic effects in a phenomenological way.
A potential that {\em cannot} produce very realistic values of the forces, stresses and energy differences 
must lack an important ingredient in its functional form
and even if it produces good values for certain physical properties (such as those to which it was fit), for other
properties it may show qualitatively different behaviour to the real system.
A good potential form should fit the {\em ab initio} data to a high degree and significantly improving this fit should 
improve the property that one is interested in simulating. If this is the case then one can be confident that the ability
to describe the property is due to a good microscopic description of the interatomic interactions.

Ultimately, our aim is to produce potentials that may be relied upon quantitatively to predict, not only the structures and
the thermodynamic properties of bulk ionic systems, but also their dynamical properties. Although
much work has been done in this direction (see, for example, references \onlinecite{madden_mgo1} and \onlinecite{madden_mgo2} and references
therein) this is still a very ambitious goal. 
Finding a potential that describes {\em dynamics} well is a particularly difficult task
\cite{angell}.

However, the fact that DFT calculations can provide an essentially limitless amount of information
that can be used in the parametrization process means that we are not constrained to using potential forms either with
a small number of parameters or with parameters that have an obvious physical interpretation.

We use the same iterative parametrization scheme that was used in reference \onlinecite{us_silica}. This is a slightly
adapted form of the schemes used in references \onlinecite{furio} and \onlinecite{laio}
and it allows us to make the very specific and non-trivial statement about a potential that for any atomic configuration
created with the potential under specified thermodynamic conditions,
the forces are, on average, within $p_{f}\%$ of those calculated {\em ab initio}\cite{abinitio},
the stress components within $p_{s}\%$, and the energy-differences between configurations
within $p_{e}\%$.

The parametrization scheme involves minimizing the 
function $\Gamma(\{\eta\}) = \omega_{F}\Delta F + \omega_{S}\Delta S + \omega_{E}\Delta E$  with respect to the
set of parameters $\{\eta\}$. 
In order to be sure that the minimization procedure is meaningful the number of {\em ab initio} configurations used in the
fit ( $n_{c}$ ) is required to be reasonably large.
In general its value depends on the system studied, the potential form used and the number of atoms in the unit cell.
It was found that for MgO a value of $n_{c}=10$ was required in order to achieve convergence in the fit $\gamma$.
In each step of the iteration at least a further five configurations were retained during the fitting procedure in order to
test that the final functional form fit these configurations as well as it did those that were used in the
minimization of $\Gamma(\{\eta\})$.

Minimization was performed using a combination of simulated annealing\cite{kirkpatrick} and
Powell minimization\cite{numerical_recipes}.
A basin in the surface defined by
$\Gamma(\{\eta\})$ in $\eta-$space was initially found using simulated annealing and, once found, further
minimization was performed using the method of Powell. Minimization of $\Gamma(\{\eta\})$using realistic force fields, and particularly simulated
annealing, is a very computationally expensive process. 
However simulated annealing is very useful for two reasons :
1. It is very stable; Powell minimization can break down if numerical errors (such as overflow errors) occur due to
unphysical values of the parameters;  2. In principle it can always bring one to the global minimum. In practice
however
this depends on how much computer time one is willing to allocate it.
These properties of the simulated annealing method make it particularly useful when fitting a potential
for the first time. One does not need to start with reasonable or physical values of the parameters in order
for it to converge and this means that one may parametrize exotic potentials for which the parameters
have no obvious physical interpretation.

The freedom that one is afforded using the combination of {\em ab initio} data and simulated annealing
is crucial. 
It simply would not be possible to parametrize a force field such as the distortable-ion
model introduced in section ~\ref{section:our_aim} without either one of these assets.
If one depends on the use of experimental data, only force-fields containing
a very small number of parameters may be parametrized if the fit is to be statistically significant. The
amount of information that can be extracted from {\em ab initio} calculations is very large by comparison and so it
allows much more complicated force-fields to be parametrized.
If parametrization of a potential depends on good initial guesses for the parameters (as is the case with most
optimization algorithms), each parameter must have an obvious physical
interpretation and this can limit ones freedom when constructing a potential form. This problem can be solved
by using simulated annealing in the parametrization process.

The {\em ab initio} calculations were performed with DFT within the local density approximation\cite{cepalder,perzung}
using the planewave pseudopotential method\cite{izc,pickett}.
We have used soft norm-conserving pseudopotentials\cite{tm} that are identical to those that have previously been used successfully to
calculate its vibrational properties for a large range of pressures and temperatures\cite{karki}. We require
our simulations to produce good quality forces and Karki {\em et al.} have performed much more rigorous tests of these
pseudopotentials than it would have been feasible for us to perform. 
We perform our calculations on unit cells containing $64$ atoms
under periodic boundary conditions and the Brillouin zone is sampled using only the $\Gamma-$point.
A kinetic energy cutoff of $120$ Ryd. was used in the plane wave expansion of the wavefunctions in order
to converge the stress very well.

\section{Many-body models for ionic systems} \label{section:manybody}
In this section a system for including many-body interactions in ionic systems will be introduced.
We confine our attention to compounds in which variations in the degree of ionicity (either locally
or globally) may be neglected. 
For such systems many body contributions to the interatomic interactions arise from the response of the size and shape of the ions
to their environment. The anions carry multipole moments and, particularly the lowest order of these, the dipole moments, 
have an important impact on phonon spectra. The change in size and shape of the anions also has an impact on the short ranged
Pauli exclusion repulsion between ions. 

Although these effects are all interdependent, it has recently been shown\cite{madden_pol,madden_compress,madden_mgo1,madden_mgo2} how
an improved description of ionic dynamics may be achieved by independently incorporating them in the interatomic
potential. 
Some of the ingredients of this force field have recently been parametrized using forces and stresses from 
density functional theory calculations and this resulted in an extremely good description of the potential energy
surface in the crystal\cite{madden_mgo2}.

The non-electrostatic repulsion between ions at short distances has in the past
usually been modelled as an exponential function of the interionic distance\cite{sangdix,bornmayer}.
The use of this exponential form rests on the assumptions that ions are spherical and
of fixed size and that the repulsion between them due to
the Pauli exclusion principle is proportional to the overlap between the ions whose charge distribution
tails off exponentially. Although this may be an adequate approximation in crystals of high symmetry at
low temperatures and a given pressure, at higher temperatures or at a different pressure or when a change of phase occurs,
it is likely that anions will readjust their size and
shape to fill the available space.

In order to cope with this, and in an attempt to improve the ability
of ionic models to reproduce experimental equations of state and the relative energetics of different crystal structures,
Wilson {\em et al.} have developed a compressible ion (CI) model\cite{madden_compress}. Previously, ionic breathing effects
had generally been incorporated within the shell model\cite{schroder}(for a recent application see reference~\onlinecite{matsui}).

Wilson {\em et al.} wrote the  potential energy due to short-range repulsive interactions between anion
and cation as
\begin{equation}
V^{\text{sr}}_{+-}(\{R_{I}\};\{\delta\sigma_{I}\}) = V_{\text{self}}(\{\delta\sigma_{I}\})
+ V_{\text{ov}}(\{R_{I}\};\{\delta\sigma_{I}\}) \label{eqn:compress}
\end{equation}
where $V_{\text{self}}$ is the sum of the changes in the internal energies of the ions and $V_{\text{ov}}$ is the
total potential energy due to the repulsive overlap interaction between the ions, and $\delta\sigma_{I}$
is the change in the radius of ion $I$ from its average value $\overline{\sigma_{I}}$.
From electronic structure calculations of the perfect cubic crystal it was found that $V_{\text{self}}$
could be written as
\begin{equation}
V_{\text{self}}(\{R_{I}\};\{\delta\sigma_{I}\}) = \sum_{I\epsilon-} D_{I}\textrm{cosh}(\beta\delta\sigma_{I})
\label{eqn:self_energy}
\end{equation}
and the standard exponential form was adopted for the interaction energy :
\begin{equation}
V_{\text{ov}}(\{R_{I}\};\{\delta\sigma_{I}\}) = \sum_{I\epsilon-,J\epsilon+}
B_{IJ}e^{-\alpha_{IJ}(R_{IJ}-(\overline{\sigma_{I}}+\delta\sigma_{I})
)}
\end{equation}
At each timestep, during a simulation, the $\{\delta\sigma_{I}\}$ were required to take values that minimized
$V^{\text{sr}}$.  

Although this has been successful in reproducing some low temperature properties of MgO and CaO,
such as the crystal energies as
a function of volume, the prediction of phonon frequencies with this model was found to be quite poor\cite{madden_compress}. This is
because the distortion of the anions are in general not spherically symmetric if the crystal is 
disordered\cite{madden_compress,madden_mgo1}.
To account for aspherical distortions of the anion, Rowley {\em et al.}\cite{madden_mgo1} have extended the previous
compressible-ion model by introducing eight further degrees of freedom to model distortions
of dipolar and quadrupolar symmetry. Self-energy functions associated with these degrees of freedom were postulated.

With the aspherical-ion (AI) model and including dispersion effects and polarization effects, Rowley {\em et al.} 
managed to find parameters that gave good phonon dispersion
curves for MgO. Very recently Aguado {\em et al.} have used an almost identical model, but with many of
the parameters found from density functional theory calculations, to produce phonon dispersion curves and thermal expansion
curves of a very high quality\cite{madden_mgo2}.
Since thermal expansion depends on the second derivatives of the potential
energy with respect to the ionic positions, this is a good test of this model's representation of the
potential energy surface of the crystal.

This method has the drawback that distortions of the anions are restricted to those of dipolar and quadrupolar symmetry. Although distortions of
different symmetry could be incorporated, this would involve a large increase in the number of degrees of freedom
and a corresponding decrease in the efficiency of the method.
It would be desirable to include the effect of distortions of arbitrary shape in a way that is practicable.

In order to implement the above model in a computationally efficient way, an extended lagrangian approach has generally been used
that is analogous to the 
Car-Parrinello method\cite{carpar} for electronic structure theory.
The degrees of freedom that describe ionic distortions ($\{\delta\sigma_{I}\}$ in the simple compressible-ion case)
are the variables with which a ``fictitious'' dynamics was associated.

There are some problems associated with the use of a Car-Parrinello approach to modelling ionic distortions however:
\begin{itemize}
\item Recent work\cite{us_carpar}
has shown that, particularly for
disordered systems, this method results in systematic errors in dynamics relative to methods in which the fictitious
variables take their minimum energy values.
\item The use of a small timestep is intrinsic to the extended lagrangian approach. For example,
in the work of Aguado {\em et al.}\cite{madden_mgo2} a step of only $0.05$ femtoseconds was used. This is more than an order of magnitude
smaller than the time step that can typically be used for non-Car-Parrinello approaches.
\item
The extended lagrangian approach is considerably more efficient during dynamics when variables are evolved from
previous time steps than it is when minimizing the energy with respect to the
extended variables for the first time. Since the parametrization process involves a large number of such minimizations,
and is computationally very expensive, the extended lagrangian approach becomes much less efficient.
\end{itemize}

\section{A Distortable Ion Model}\label{section:our_aim}
Our goal is to develop a general mathematical framework within which the many-body effects
of ionic compression and aspherical ionic distortion can be modelled, but that
avoids the problems associated with the previously proposed models.
In particular we would like to find a computationally efficient way of modelling
these effects that avoids the use of an extended lagrangian and the introduction of 
fictitious degrees of freedom. 

There are many open questions regarding the interactions between simple (electronically isolated) ions
and so the framework should be general enough to
allow for continual improvement of the specific form of the potential.
For example, the form of the self-energy in equation~\ref{eqn:self_energy} and the self energies associated with aspherical 
distortions can certainly be improved. It may also be worth investigating different forms of the overlap energy.

We will be primarily concerned with the anion-cation interaction. Of much lesser concern, initially at least,
is the anion-anion interaction energy that has been found to provide only $\sim3\%$ of the
energetics of the perfect
crystal\cite{madden_mgo1,madden_compress}.
Although, the same cannot be said with any degree of certainty of more disordered
phases, or systems of different stoichiometry such as SiO$_{2}$, it is nonetheless
the most
obvious place to start when constructing a potential.

We assume that a distortable ion ( such as O$^{2-}$ ) has its shape and size ``influenced'' by all sufficiently
close neighbouring ions.
Much as in the scheme of Wilson, Madden and coworkers\cite{madden_compress,madden_mgo1}, an ion is described as a nucleus surrounded by a
single membrane (representing the electrons) the radius
of which is allowed to vary with the two polar angles (although in their case,
the radius only varied in certain
symmetric ways).
The influence an ion $J$ exerts on ion $I$  can loosely be
thought of as a restraining force on the ion's tendency to expand and this restraint has a dependence
on the polar angles  $(\theta,\phi)$ in the spherical coordinate system centered on $I$.
We also assume that the influence
exerted at coordinates $(\theta,\phi)$ is zero if the angle between the outward unit vector at those coordinates
${\mathbf \ell}(\theta,\phi)$ and
the vector ${\mathbf R}_{JI} = {\mathbf R}_{J}-{\mathbf R}_{I}$ is greater than $90^{\circ}$.

We write the total influence on $I$ at $(\theta,\phi)$
due to all the other ions as
\be
\rho_{I}^{(1)}(\theta,\phi) = \sum_{J\neq I} f_{IJ}(R_{IJ}){\mathbf \ell}(\theta,\phi)\cdot {\mathbf x}_{JI}
\Theta( {\mathbf \ell}(\theta,\phi)\cdot {\mathbf x}_{JI}) \label{eqn:Theta}
\ee
where ${\mathbf x}_{JI}={\mathbf R}_{JI}/R_{JI}$ and
\begin{equation}
\Theta( {\mathbf \ell}(\theta,\phi)\cdot {\mathbf x}_{JI}) =
1 \;\;\text{if} \;\;{\mathbf \ell}(\theta,\phi)\cdot {\mathbf x}_{JI} > 0 \\
\;\;\;\text{and}\;\;\;
0 \;\;\text{if} \;\;{\mathbf \ell}(\theta,\phi)\cdot {\mathbf x}_{JI} < 0
\end{equation}
Functional forms for $f_{IJ}(R)$ will be discussed in Section IV.A.  The subscripts $IJ$ are to indicate 
that a different function is used for each distinct pair of ionic species.

Apart from a multiplicative constant, the spherical average of $\rho_{I}^{(1)}(\theta,\phi)$ is
\be
\rho_{I}^{(0)} = \sum_{J\neq I} f_{IJ}(R_{IJ}) \label{eqn:rhoi0}
\ee
We write the angular dependent radius
, $\sigma_{I}(\theta,\phi)$ of an ion as
\be
\sigma_{I}(\theta,\phi) = \sigma_{I}^{(0)}(\rho^{(0)})
+ \sigma_{I}^{(1)}(\rho^{(0)},\rho_{I}^{(1)}(\theta,\phi))
\ee
In other words, the radius at $(\theta,\phi)$ is written as a sum of an average value due to the
influence of all the ions and deviations from that average.
Functional forms for $\sigma_{I}^{(0)}$ and $\sigma_{I}^{(1)}$ will be discussed in the next section.
The distance between the membranes of ions $I$ and $J$ along their line of centers is
\be
L_{IJ} = R_{IJ} - \sigma_{I}(\theta_{JI},\phi_{JI}) - \sigma_{J}(\theta_{IJ},\phi_{IJ}) \label{eqn:L}
\ee
where $\theta_{JI}$ and $\phi_{JI}$ are defined such that ${\mathbf \ell}(\theta_{JI},\phi_{JI})={\mathbf x}_{JI}$.
We will use the notation
\begin{eqnarray}
\rho_{IJ}^{(1)} &=&  \rho_{I}^{(1)}(\theta_{JI},\phi_{JI}) \\
\sigma_{IJ}^{(1)} &=&  \sigma_{I}^{(1)}(\rho^{(0)},\rho_{I}^{(1)}(\theta_{JI},\phi_{JI}))\\
\sigma_{IJ} &=&  \sigma_{I}(\theta_{JI},\phi_{JI}) 
\end{eqnarray}

We now define the contribution to the total energy of the system from the short-range repulsive interactions
as a sum of pairwise interactions between membranes.
\be
U^{\text{\tiny SR}} = \sum_{I,J>I}U^{\text{\tiny SR}}_{IJ}(L_{IJ})g_{IJ}(R_{IJ}) \label{eqn:energy_aim}
\ee
where $g_{IJ}(R)$ takes the value $1$ for $R<R_{a}$, $0$ for $R>R_{b}$ and
decays smoothly from $1$ to $0$ between $R_{a}$ and $R_{b}$. This allows us to truncate the
interaction at intermediate distances.

The expression for the forces on the ions is given in Appendix A.

\subsection{Applying the model}\label{section:applying}
In order to apply this model we clearly need to find suitable expressions for the functions
$f_{IJ}$,$\sigma^{(0)}_{I}$, and $\sigma^{(1)}_{IJ}$.
We begin by making the assumption that
the most important interaction is the anion-cation interaction although this will be extended
at a later stage to include the anion-anion interaction in a limited way.
For the moment we are concerned with systems with two species such as MgO and we assume that the
cation is small and rigid. For MgO it is likely that this is a very good assumption, given its
degree of ionicity.

In order to draw a correspondence with the compressible ion model of Wilson {\em et al.}\cite{madden_compress}
(see section ~\ref{section:manybody}) we write the total energy of the system due to short-range
repulsion as
\begin{eqnarray}
V^{\text{SR}} &=& \sum_{I\epsilon-} V_{I}^{\text{self}}(\sigma_{I}^{(0)})+
\sum_{I\epsilon-,J\epsilon+}B_{+-}e^{-\alpha_{-+}(R_{IJ}-\sigma^{(0)}_{I})}\nonumber \\
&+&\sum_{I,J\epsilon-,J>I}B_{--}e^{-\alpha_{--}R_{IJ}}
 + \sum_{I,J\epsilon+,J>I}B_{++}e^{-\alpha_{++}R_{IJ}}
\end{eqnarray}
The values of the anion radii at any time should be such that this repulsive energy
is minimized. In other words
\begin{eqnarray}
\frac{\p V^{\text{SR}}}{\p \sigma_{I}^{(0)}} &=& 0 \;\;\;,\;\;\; \forall I  \\
\Rightarrow \frac{\p V_{I}^{\text{self}}}{\p \sigma_{I}^{(0)}} + \alpha_{-+}e^{\alpha_{-+}\sigma_{I}^{(0)}}
\sum_{J\epsilon+}B_{+-}e^{-\alpha_{-+}R_{IJ}}&=&0
\end{eqnarray}
To simplify the notation we write $B' =  \alpha_{-+}B_{+-}$ and
$\zeta(\sigma_{I}^{(0)})=\frac{\p V^{\text{self}}}{\p \sigma_{I}^{(0)}}$.
\be
\zeta(\sigma_{I}^{(0)})e^{-\alpha_{-+}\sigma_{I}^{(0)}} = - \sum_{J\epsilon+}B'e^{-\alpha_{-+}R_{IJ}}
\label{eqn:zeta}
\ee

As was noted previously in section~\ref{section:our_aim}  there has been some discussion about the form of the self-energy of compressible
ions. Although in the paper by Wilson {\em et al.}\cite{madden_compress} the form used was that of a hyperbolic cosine
of the amount of compression $\delta\sigma$, applications of the shell-model generally use a harmonic 
expression for this energy \cite{matsui,schroder} and in a very
recent paper \cite{finnis} Marks {\em et al.} have argued that for the oxide ion one should treat
the $2p^{6}$ shell and the $s^{2}$ shells separately with harmonic and exponential compression energies
respectively.
In references \onlinecite{madden_compress} and \onlinecite{finnis} justification of the forms of $V_{\text{self}}$ 
used were based on quantum chemical calculations of the cold crystal under a large range pressures.
The self-energy of ions in disordered systems may be quite different to that in the perfect crystal and
anyway, for interatomic distances that one should expect to find in the crystal at zero pressure and temperatures up to the melting point,
the calculated self-energy is still close to a linear regime. Furthermore we do not see any compelling physical
reasoning behind any of the forms used. For these reasons we consider the form of this energy to be an open question.
As a preliminary test of our model we have chosen an exponential form for $V_{\text{self}}$  as this simplifies
the mathematics considerably.
$\zeta(\sigma_{I}^{(0)})$ will then also have an exponential form so we may write
\be
A_{I}e^{-\beta_{I}\sigma_{I}^{(0)}}e^{-\alpha_{-+}\sigma_{I}^{(0)}} = - \sum_{J\epsilon+}B'e^{-\alpha_{-+}R_{IJ}}
\ee

By merging constant terms to simplify the notation, this equation can be rewritten in the form
\be
\sigma_{I}^{(0)}(\rho_{I}^{(0)}) = C_{1} + C_{2}\ln(\rho_{I}^{(0)}) \label{eqn:sigma0}
\ee
where we say that
\be
\rho_{I}^{(0)}=\sum_{J}C_{3}e^{-C_{4}R_{IJ}}
\ee
and $C_1$,$C_2$..etc are constants.
By analogy with equation~\ref{eqn:rhoi0} we can say that
\be
f_{IJ}=C_{3}e^{-C_{4}R_{IJ}} \label{eqn:fij}
\ee
One is not confined to such simple forms for the self-energy but for many forms one cannot write
equation~\ref{eqn:zeta}
in terms of $\sigma_{I}^{(0)}$ and one is forced to find $\sigma_{I}^{(0)}$ by an iterative procedure.
This only has a very slight impact on the efficiency of calculating the potential.
Another form that we have tried, and for which this procedure is used is
\be
V_{I}^{\text{self}}(\sigma_{I}^{(0)}) = \frac{\epsilon_1}{\epsilon_2 + \sigma_{I}^{(0)}}
+ \frac{\epsilon_3}{(\epsilon_4 + \sigma_{I}^{(0)})^2} \label{eqn:selfenergy2}
\ee
where $\epsilon_1$,$\epsilon_2$..etc are constants.
This form was chosen according to the (admittedly, highly simplistic) physical reasoning that the internal factors
that determine an ion's radius are the electrostatic energy which varies like the inverse of a distance
and the kinetic energy of the electrons which varies like the inverse of a distance squared.

 The above analysis has shown that the distortable-ion model presented is mathematically equivalent to the
compressible-ion model of Wilson {\em et al.} if $\sigma_{IJ}^{(1)}=0$ in the limit that the
fictitious mass of the extended-lagrangian approach goes to zero.

It is not possible to map our approach onto the aspherical-ion model.
However, we take a different approach to aspherical distortions.
Given the functions $f_{IJ}$ and $\sigma^{(0)}_{I}$ as a starting point we may postulate
a form for $\sigma_{IJ}^{(1)}$.
We assume that the local distortion of an ion's membrane scales with the local ''density''
$\rho^{(1)}$ in the same way as the average radius scales with the spherical average of the 
density $\rho^{(0)}$.
We therefore write 
\be
\sigma_{IJ}^{(1)} = C_{5}\ln\bigg(\frac{\rho_{IJ}^{(1)}}{\rho_{I}^{(0)}}\bigg)
\label{eqn:sigma1}
\ee
Since we will be parametrizing this force by fitting to {\em ab initio} data, the minimization 
routine has the freedom either to make the constant $C_5$ very small or zero if this
is not a reasonable functional form, or if aspherical distortions are really energetically
equivalent to spherical ones it can make $C_2=C_5$ in which case the distortions
of purely spherical symmetry disappear and
\be
\sigma_{IJ} = C_1 + C_{5}\ln(\rho_{IJ}^{(1)})
\ee

Although all the above derivation has assumed that this potential is only to be used for modelling cation-anion
interactions, the generality and freedom afforded us by our parametrization procedure 
means that we lose nothing by trying to apply it to the anion-anion
interaction also. We have done this by fitting parameters for the anion-anion interaction and we have
found that it does improve the ability of the model to fit the {\em ab initio} forces.
A more sensible, but also more expensive way of tackling the anion-anion interaction would be to
introduce a self-consistent procedure to minimize the angular dependent radii simultaneously.

We also note that, as has been pointed out by Marks {\em et al.}, different electronic shells
have different compression characteristics. This could be modelled within the present scheme by having two
or more such distortable ion potentials acting in parallel.  

In our application of this model we have generally used timesteps of between $1$ fs 
and $1.5$ fs and the model has been found to be extremely efficient. 
As an example: in a simulation of $512$ atoms of crystalline MgO at $3300$ K 
(using potential F which is discussed in section~\ref{section:altering}) we have used a time step of $1$ fs.
In this simulation, the time required for each calculation of the distortable-ion contribution to energy, forces, and stress 
was $0.57$ seconds on a single $300$Mhz SGI origin MIPS R12000 processor. 
On the other hand, the contribution of all the electrostatic terms was $3.0$ seconds.
The total energy in this simulation drifted by approximately $2$ K per picosecond during this simulation. This energy drift
can be almost eliminated by more fully converging the polarization at each time step.

A point to note regarding equation ~\ref{eqn:Theta} is that there is a discontinuous change in $\Theta$
when ${\mathbf \ell}(\theta,\phi)\cdot {\mathbf x}_{JI} = 0$. Although equation ~\ref{eqn:Theta} itself is
not discontinuous, its derivative with respect to the ionic positions is. This leads to discontinuous
forces and consequently to a drift in the total energy. In practice we have not found this to be a problem
as this drift in energy is very small compared to the drift that is due to the incomplete convergence of the
polarization. However, if necessary this problem may be eliminated by replacing $\Theta$ with a function that 
varies smoothly from $1$ to $0$ as ${\mathbf \ell}\cdot {\mathbf x}_{JI}$ approaches zero.

\subsection{Testing the model}
The model that we have proposed is considerably more complex than a pair potential and so
in testing the model we first would like to verify that it improves upon simpler force fields.
We confine ourselves to testing the usefulness of three different ingredients of the interionic potential:
\begin{enumerate}
\item
A pairwise short-range interaction potential of the form
\begin{equation}
U^{\text{\tiny SR}}_{IJ}(R_{IJ}) = B_{IJ}e^{-\alpha_{IJ}R_{IJ}}
-\frac{C_{IJ}}{R_{IJ}^{6}}-\frac{E_{IJ}}{R_{IJ}^{N_{IJ}}}
\end{equation}
where $B_{IJ},\alpha_{IJ},C_{IJ},E_{IJ},N_{IJ}$ are all parameters to be optimized. 
\item
A polarizable-ion potential including short-range polarization, as discussed in reference~\onlinecite{us_silica}.
Only the oxygen ion is considered polarizable.

\item
A distortable-ion potential, as discussed in sections~\ref{section:our_aim} and ~\ref{section:applying}.
The interaction energy between ions $I$ and $J$ is given by
\be
U^{\text{\tiny SR}}_{IJ}(L_{IJ}) = A_{IJ}e^{-\alpha_{IJ}L_{IJ}} + B_{IJ}e^{-\beta_{IJ}L_{IJ}} \label{eqn:ab_def}
\ee
and the functions $f_{IJ}$,$\sigma^{(0)}_{I}$, and $\sigma^{(1)}_{IJ}$ are given the same forms as
in equations~\ref{eqn:fij},~\ref{eqn:sigma0}
and~\ref{eqn:sigma1} respectively, i.e.
\begin{eqnarray}
f_{IJ}=C^{(1)}_{IJ}e^{-C^{(2)}_{IJ}R_{IJ}} \label{eqn:fij2} \\
\sigma_{I}^{(0)}(\rho_{I}^{(0)}) = C^{(3)}_{I}\ln(\rho_{I}^{(0)}) \label{eqn:sigmai} \\
\sigma_{IJ}^{(1)} = C^{(4)}_{IJ}\ln\bigg(\frac{\rho_{IJ}^{(1)}}{\rho_{I}^{(0)}}\bigg) \label{eqn:sigmaij}
\end{eqnarray}
where $C_{1}$ from equation~\ref{eqn:sigma1} has been merged into the pre-exponential factors $A_{IJ}$ and $B_{I
J}$.
The parameters to be optimized are $A_{IJ}$, $\alpha_{IJ}$, $B_{IJ}$, $\beta_{IJ}$, $C^{(1)}_{IJ}$, $C^{(2)}_{IJ
}$, $C^{(3)}_{I}$
,$C^{(4)}_{IJ}$.
The values $R_{a} = 8.5$ a.u and $R_{b} = 10$ a.u. were used in the decay function $g_{IJ}$.
\end{enumerate}
In addition to the ingredients mentioned, each force field also included the point-charge electrostatic
potential with the charge on an ion as a parameter.

We have parametrized five different force fields, as follows:
\renewcommand{\theenumi}{\Alph{enumi}}
\begin{enumerate}
\item
A pair-potential : short-range pair potential, parametrized in the crystal at ambient conditions.
\item
A polarizable potential :  short-range pair potential, with polarizable anions, parametrized in the crystal at 
ambient conditions.
\item
A distortable-ion potential :  distortable-ion potential, parametrized in the crystal at ambient conditions.
\item
The full model :  distortable-ion potential, with polarizable anions, parametrized in the crystal at ambient conditions.
\item
The full model : distortable-ion potential, with polarizable anions, parametrized in the liquid at $3000$ K.
\end{enumerate}

In order to avoid the computational expense of performing a full self-consistent parametrization procedure for each
of these potentials we have only used this procedure to parametrize three potentials using the full model (distortable-ion model
with point charges and dipole polarization). The three potentials were optimised at zero pressure for (i)
the liquid at $3000$ K (ii) the crystal at $2000$ K and (iii) the crystal at $300$ K respectively. Each of these 
three potentials was used to create trajectories at the conditions for which it was optimised from which
'snapshot' atomic configurations were extracted and used in {\em ab initio} calculations. We will show that
the full model is the best at reproducing the {\em ab initio} data and so these configurations were considered to 
be as realistic as we have the ability to create.

Each of the potentials A to E was then parametrized using {\em ab initio} data from $10$ of these configurations (at the relevant conditions).
Each potential thus created was then tested on its ability to fit $10$ new configurations (i.e. that were not used in the parametrization process) 
and also $10$ configurations at each of the other two sets of conditions. For example. potential A was parametrized at $300$ K and then its
ability to fit the {\em ab initio} data in the crystal at $2000$ K and the liquid at $3000$ K was also tested.
In all cases the error in the stress was evaluated relative to a pressure of
$B=140$ GPa.

The results are summarized in table~\ref{table:pot_test}. We cannot guarantee that we have found the global minimum
in each case during optimization as simulated annealing had to be done at a rather rapid quench rate. The simulated annealing
was followed by Powell minimization\cite{numerical_recipes}.
In each case, the total minimization time was the same ($10$ days on a single processor) and
therefore more economical force-fields are likely to be better minimized than less economical ones.
A number of things can clearly be seen from table~\ref{table:pot_test}.
 First of all, not surprisingly, the distortable-ion model on its own is quite
bad. This is probably because of the shortness of the range of its interactions. Ions further away from each other than $10$ a.u.
interact only via the coulomb force between their charges.
At $300$ K, the full model is clearly better than all other forms. It also transfers very well
up to higher temperatures and to the liquid. The pair-potential, although working quite
well for the crystal, does not transfer well to the liquid. The polarizable model yields results that are
intermediate in quality
between the pair-potential and the full model.
The results are a clear illustration of the fact that by adding more physics into the form of an effective potential one can create
force-fields with, not only an improved ability to fit the {\em ab initio} data, but also a much improved transferability
between different phases and conditions.

The poor fit of potential E to the energy differences in the crystal at ambient conditions is because the energy
differences in the liquid and high temperature solid are much greater than those at lower temperatures. The
absolute value of the error in the energy differences is the same at low temperature and high temperature but $\Delta E$
is the error relative to the root-mean-squared value, which for the crystal is very small.

\subsubsection{Phonon Frequencies}
Having established that our inclusion of many-body effects has improved the potential form with respect to the pair potential,
at least according to the criterion that we have adopted,
we now look at its ability to model the vibrational spectrum of MgO.
We note once again that the DFT scheme to which the potential was fit gives a very good description of phonon
frequencies at ambient conditions \cite{karki}.

We find the phonon frequencies for our potential at a number of k-points from the positions of the peaks in the
spectra of the spatial Fourier components of longitudinal and transverse charge and mass current correlation
functions\cite{quadrupole}.

We performed an MD simulation on a system of $512$ atoms
using the full-model, optimised in the crystal at $300$ K (potential D). The current correlation functions were calculated
on a time domain of length $2.9$ ps that was averaged over a simulation of length $20$ ps.
The phonon dispersions that we get are shown in
figure~\ref{fig:phonon_full}. We get an extremely close fit to both the experimental and the self-consistent DFT
data. The chief discrepancies are in the optical modes which are systematically underestimated. The longitunical optical
mode in particular is underestimated near the zone center. Although we do not calculate the mode frequencies at
$\Gamma=(0,0,0)$, as this would require an infinitely large simulation cell with the method that we are using,
it looks as though the splitting between the longitudinal optical (LO) and the transverse optical (TO) phonons is slightly underestimated.
In our parametrization procedure we have used a small cell to perform the {\em ab initio} calculations and so the
long-range interactions that are important for dispersion near the $\Gamma-$point are not included. Our hope is that
by modelling correctly the electrostatics at shorter range, we get a potential that, when used in a larger simulation
cell, can accurately model the long range electrostatic interactions. This is not guaranteed however and is likely
to work only if we include all relevant screening mechanisms in our functional form.
The incorrect LO-TO phonon splitting suggests that our
description of the electrostatics is incomplete since it arises from the long-range electric field induced by the 
LO phonon.
This is not surprising since dipole polarization is only one of many
screening mechanisms that are present in the real system. 
It may be that charge-transfer between ions is important. However, a comparison with the results
of reference~\onlinecite{madden_mgo2} is suggestive of it being due to the fact that we haven't included the affects
of higher-order multipoles.
Density functional perturbation theory calculations\cite{karki} of the Born effective charges
yield values of $Z_{B}^{*}=\pm 1.94$.
However,
the charge on the oxygen ion in this potential (and all other potentials
that we have fit) is $\sim 1.5$ - considerably less than this.
Under the assumption that, within our model, short-range interactions and electrostatics
describe completely separate aspects of the potential-energy surface (we do not know the extent to which this is true)
the minimization routine fits the
charge and the polarizability so as best to approximate the electrostatics of the crystal.
The lack of higher order multipoles means that it must choose a compromise between purely dipole screening, in which
the polarizability $\alpha$ and the charge $q$ take their ``true'' values, and uniform screening in which the charge
is simply reduced by a factor equal to the dielectric constant and the polarizability is zero.
In reference~\onlinecite{madden_mgo2} they use formal ionic charges and include both quadrupoles and dipoles and
they get better agreement with experiment. Nevertheless, the description of the electrostatics that we have is
significantly better than any other effective potential that we are aware of
and quadrupoles would add considerably to the computational expense of the model.

It is also worth noting that, although in reference~\onlinecite{madden_mgo2} the potential used 
seems to give a slightly
better description of the optical phonon frequencies, they report a fit to the DFT forces of
$\Delta F < 10 \%$. Potential D is in significantly better agreement with our DFT data than this. 
Direct comparison of these two fits is not entirely justified, however. 
Aguado {\em et al.} fit to {\em ab initio} calculations of three different 
crystalline phases and therefore to three different
coordination environments. Our fit of potential D is restricted to the phase and conditions at which
we calculate the phonon frequencies. Although our fit is converged with respect to the
number of atomic configurations tested, the small size of the fitting cell and the low temperature 
and high symmetry of these fitting conditions
mean that only a very small region of phase space is visited. It may be that our functional form 
can numerically reproduce the potential energy surface in this small region of phase space in
a way that is not necessarily physical and that therefore does not extend very well to describe
the longer wavelength phonons that are present when we move to the larger simulation cell
used for the calculation of phonon frequencies. 

On the other hand, if one is to construct a potential for use in a given region of phase space
it may not always be advisable to include configurations from regions of phase space that are too far away
from this domain of application.
There is a danger that changes in electronic structure that cannot be represented phenomenologically
by the functional form of the potential may occur. For example, a change in coordination of the ions may
be accompanied by a change in the degree of ionicity. Since our model keeps the charges on each ion fixed, 
by fitting to different coordination environments we may find charges that are the best compromise
between the optimal values for each different environment, and that are therefore ideal for
none of them. For many materials, the liquid structure is not too dissimilar to the solid at the same
pressure and temperature and may therefore be a good compromise as a fitting environment. By fitting
to the liquid we can visit a large volume of phase space that is nevertheless centered on or near the
crystalline phase of interest. 

We have used the above argument simply as a transparent way of discussing issues to consider when
parametrizing a potential. The success of the potential created by Aguado {\em et al.} 
suggests that, for MgO, changes of ionicity do not present a problem in practice.

\subsection{Altering the model}\label{section:altering}
The simple exponential form that we have used for the self energy was chosen for its simplicity. 
We would like to test other forms of this energy and although we do not go in detail into this problem here,
we do parametrize a potential using the form of equation~\ref{eqn:selfenergy2}. This has an appealing
physical form and it requires one to self-consistently find the values of the radii $\sigma^{(0)}_{I}$ and
$\sigma^{(1)}_{IJ}$. 
We observed for the previously created potentials that changing the radial cutoffs for the decay function
$g_{IJ}$ of the distortable
ion model, $R_{a}$ and $R_{b}$, did not significantly change the fit to the {\em ab initio}. 
For larger distances the forces involved are either too small or cancel one another out.
For this reason, and in order to improve efficiency we have used the slightly smaller
values of $R_{a}=7.0$ a.u. and $R_{b}=8.0$ a.u.

Using the full model we perform a full self consistent parametrization in the liquid
at $3000$ K (potential F) and the solid at $3000$  K (potential G).
We have consistently found during our parametrizations that no significant 
improvement to the fit is obtained by including dispersion terms. It is likely that 
this is, at least partly, related to the limitations of the local density approximation as there
is no reason to believe that it can describe such highly non-local dynamical fluctuations.
Nevertheless, in the parametrization of potential F, $1/R^{6}$ and $1/R^{8}$ 
terms were used in conjunction with Tang-Toennies dispersion damping functions\cite{tangtoennies}. 
The fact that the parametrization procedure made some of these interactions
repulsive (see table~\ref{table:parameters_disp}) is a clear indication that they should 
not necessarily be interpreted as representing the effect of electron dispersion.

We have then tested the ability of potentials F and G to fit {\em ab initio} data
at these conditions. For this testing we use $20$ solid configurations and $20$ liquid
configurations. The results are summmarized in table~\ref{table:pot_test2}.
The fit is extremely good and even better than that obtained with the previous form of the self-energy.
However, a problem has been encountered with the potential that was parametrized on the solid at $3000$ K (potential G).
We found that at higher temperatures and in the liquid, the iterative procedure by which the radii ($\sigma_{IJ}$) were
found sometimes failed to converge. This is the explanation for the poor fit of potential G to the liquid data.
This highlights the importance of caution when applying a force field to conditions
different from those in which it was parametrized.
During extensive simulations of the liquid the distortable ion model for the potential parametrized on the liquid 
never failed to converge.

The phonon dispersions for potential G are shown in figure~\ref{fig:phononF}.
One should not expect results that are as good as those for
a potential that is parametrized at ambient conditions, and so the results are extremely good. There is very good
agreement with both experiment and the DFPT results of Karki {\em et al.}. As before, the worst agreement
is for the long-wavelength LO phonons, and once again this is probably due to our incomplete description
of electronic screening. It may also be that the very high symmetry of the relatively cold crystal makes
polarization energetically unfavourable, and so the polarizability appropriate for a hot crystal is larger.
A too-large polarizability should manifest itself in the phonon curves as a
lowering of the energy of the long-wavelength LO phonon modes due to improved screening of the macroscopic electric field.
Nevertheless, in general the results seem even better than those of figure~\ref{fig:phonon_full} and the ability
of both potentials to reproduce {\em ab initio} energy differences is very satisfying and suggests that
the form of the distortable-ion self-energy used may be better than a simple exponential.

\subsubsection{Density}
Figure~\ref{fig:eos} shows the equation of state of crystalline MgO at $300$ K for potentials
D and G. 
Although both are in extremely good agreement with experiment at low pressures, potential
D is better at much higher pressures. There are a number of possible reasons for this. First of all,
although we should not expect our DFT calculations to produce exactly the same equation of state as that
of Karki {\em et al}\cite{karki} due to differences in the details of the calculations\cite{eos}, their
equation of state does become increasingly inaccurate at high pressures. It may be that the
improved fit of potential G to the {\em ab initio} data results in a disimprovement in the equation
of state due to an inadequacy of the {\em ab initio} data.
Another possibility is that the reduced values of $R_{a}$ and $R_{b}$ result in a reduction of transferability
to high pressures due to the gradual introduction of more shells of neighbours in the distortable-ion calculation.

Figure~\ref{fig:thermal} shows the density as a function of temperature for potentials F and G (Potential D gave very
similar results to potential G for a system size of $512$ atoms). The experimentally observed 
thermal expansion is clearly very well reproduced by our potential. What is striking is that, particularly at low temperatures,
finite size effects are very small. In reference \onlinecite{madden_mgo2} finite size effects are much bigger. 
It may be that our model benefits from fitting the electrostatic interactions to the DFT data so that screening of interactions
is more effective. It may also be that our potential {\em over}-screens the electrostatic interactions, i.e. that the dipole
polarization is too large. This would be consistent with our underestimation of the LO mode frequencies approaching
the $\Gamma$-point.

Our potentials overall give an extremely good description of the density as a function of pressure and temperature.

\section{Analysing the model}
As discussed in section~\ref{section:applying}, we do not impose the distortable-ion model on the
system. We have parametrized the force-field using simulated annealing that was begun at a high temperature. This means
that, although we have supplied a functional form that is capable of including distortable-ion
behaviour, the minimization routine is free to do with this form whatever is best for reproducing {\em ab initio}
forces. The options that are open to the minimization routine are
\begin{itemize}
\item to disable all variable-radius functionality, and therefore to model the interionic forces
with a double exponential of the interionic distance $R_{IJ}$.
It would be optimal to do this if the way in which we model distortions
is completely unphysical.
\item to enable only the compressible-ion part of the model, i.e. that which is analogous to the
model of Wilson {\em et al}\cite{madden_compress}, thereby allowing only spherically symmetric anion distortions.
It would be optimal for it to do this if the way in which we model aspherical distortions is unphysical but our description of 
spherical distortions is reasonable.
\item to enable only the asymmetric part of the model and to disable purely spherically-symmetric distortions.
This is optimal if our reasoning that aspherical distortions are energetically equivalent to spherical ones is true and
the form of the model is reasonable.
\item to partially enable either or both types of distortions as the best compromise between
rigid-ion behaviour, breathing-ion behaviour and distortable-ion behaviour if all three of the
models fail to varying degrees and in different ways to reproduce the {\em ab initio} potential energy surface.
\end{itemize}
The parametrization process is therefore itself a test of the distortable-ion model.
We now look at what, precisely this parametrization process has done by examining the radius
of an oxygen ion in the direction of a neighbouring magnesium ion for one of our potentials (potential F
that is discussed in section~\ref{section:altering}).
The test is performed in the crystal at $3000$K.
The local radii of the anions consist of an arbitrary constant, that may be merged into the
constant coefficient of the exponential force between ions, and the true variations of the radii
due to changing environment.  We look at the quantities
$\sigma_{IJ}-\overline{\sigma_{IJ}}$
, $\sigma^{(1)}_{IJ}-\overline{\sigma^{(1)}_{IJ}}$,
and $\sigma^{(0)}_{I}-\overline{\sigma^{(0)}_{I}}$ for anion $I$ and cation $J$ where
$\overline{\sigma_{IJ}}$,$\overline{\sigma^{(1)}_{IJ}}$ and $\overline{\sigma^{(0)}_{I}}$ are averages
over a long trajectory.
These quantities are therefore the non-constant parts of the different contributions to the radius
of ion $I$ in the direction of $J$ (Recall that $\sigma_{IJ} = \sigma^{(0)}_{I} + \sigma^{(1)}_{IJ}$ where
$\sigma^{(0)}_{I}$ includes only spherically-symmetric distortions and $\sigma^{(1)}_{IJ}$ includes
aspherical distortions ).

The results are shown in figure~\ref{fig:radius} and the variation
in the value of $L_{IJ}$, as defined by equation~\ref{eqn:L}, along the same
trajectory is shown for comparison.
As can be seen, the local radius is dominated by the effect of the {\em aspherical} part of the
distortable-ion model. The spherical part makes a significantly smaller contribution. This
clearly vindicates our extension of the compressible-ion model to include aspherical distortions.

The variation in the radius is very small compared to the variation in $L_{IJ}$ and so we look at
what contribution this makes to the forces between the ions.
Looking at the forces in a pairwise way is not entirely justified given the many-body nature
of the potential, however it seems natural to look at the quantities
\be
Q^{(1)}_{IJ} = 100\times\frac{
           \frac{
            \partial U^{\text{\tiny SR}}_{IJ}(L_{IJ})
           }
           {
            \partial L_{IJ}
           }
         - \frac{
            \partial U^{\text{\tiny SR}}_{IJ}(R_{IJ}-\overline{\sigma_{IJ}}- \overline{\sigma_{JI}})
           }
           {
            \partial R_{IJ}
           }
         }
         {
          F^{\text{r.m.s.}}_{IJ}
         } \label{eqn:q1}
\ee
and
\be
Q^{(2)}_{IJ} = 100\times\frac{
           \frac{
            \partial U^{\text{\tiny SR}}_{IJ}(L_{IJ})
           }
           {
            \partial L_{IJ}
           }
         - \frac{
            \partial U^{\text{\tiny SR}}_{IJ}(R_{IJ}-\overline{\sigma_{IJ}}- \overline{\sigma_{JI}})
           }
           {
            \partial R_{IJ}
           }
         }
         {
          \frac{\partial U^{\text{\tiny SR}}_{IJ}(L_{IJ})}{ \partial L_{IJ}}
         } \label{eqn:q2}
\ee
where $F^{\text{r.m.s.}}_{IJ}$ is the root mean-squared value (averaged over time)
of the {\em total} force on anion $I$ (i.e. from all atoms
and from both electrostatic and non-electrostatic contributions) projected onto the
line joining the centers of $I$ and $J$. These quantities are plotted in figure~\ref{fig:force}.
$Q^{(1)}$ is a way of looking at the impact of instantaneous variations of the membrane radii on the
total force on the ion.
$Q^{(2)}$ is a way of looking at the impact of instantaneous variations of the membrane radii
on just the {\em short-range} part of the force between ions $I$ and $J$.
If the radius of the ion is constant, then $Q^{(1)}=Q^{(2)} = 0$.
It is difficult to know how one should best compare forces, or judge the
impact of individual contributions to the forces.
However, inspection of these two quantities strongly suggests that, with the
parameters of the model chosen by the minimization routine, the
variation of the anion's radius has a significant impact on dynamics.

The above discussion shows that the minimization routine finds it optimal to allow fully aspherical
distortions of the anions that impact significantly on the interatomic forces.

\section{Discussion}
In this paper we have presented a many-body interatomic potential for ionic systems that attempts
to model the effect on interionic interactions of ions breathing and distorting as their environments change.
The potential has been used in conjunction with a rigorous {\em ab initio} parametrization
scheme and applied to the case of bulk magnesium oxide.
The only empiricism involved in our construction of these potentials has been in the
choice of the form of the potential. We have justified this form by its ability to
fit the density functional theory potential energy surface and to reproduce experimental data.

We have clearly shown that the form of the potential presented has a significantly improved ability (with respect to
pairwise interactions) to fit the
ionic potential energy surface of the hot crystal and the liquid calculated within density functional
theory.  
It also has an improved transferability between phases and different physical conditions.

Once parametrized carefully, the potential has been shown to produce excellent phonon dispersion curves,
equations of state and thermal expansion. 

The computational expense of this potential is very small compared to other ingredients of {\em realistic}
potentials such as dipole polarization. It is also likely to be much less expensive than previously proposed
extended lagrangian force-fields\cite{madden_compress,madden_mgo1}, particularly since it allows the
use of much larger timesteps.

For all these reasons, the potential proposed is a valuable addition to effective force fields that aim towards a quantitative
description of real ionic materials.

The mathematical form of the potential is highly amenable to improvement and research into the 
optimal forms of its constituent functions may be expected to improve its ability to fit {\em ab initio} data.

\begin{section}*{Appendix A}
Here we derive the expression for the forces on the ions from the definition
of energy given in section~\ref{section:our_aim}

Using equations~\ref{eqn:L} and~\ref{eqn:energy_aim} the $\alpha$th force component on ion $K$ 
may be written as
\begin{eqnarray}
F_{K}^{\alpha} = &-&\sum_{I,J>I}g_{IJ}\frac{\partial U^{\text{\tiny SR}}_{IJ}}{\partial L_{IJ}}\bigg (x_{IJ}^{\alpha}
(\delta_{IK}-\delta_{JK}) -
\frac{\partial \sigma_{IJ}}{\partial R_{K}^{\alpha}}
- \frac{\partial \sigma_{JI}}{\partial R_{K}^{\alpha}} \bigg )\nonumber \\
&-& \sum_{I,J>I}U^{\text{\tiny SR}}_{IJ}\frac{\p g_{IJ}}{\p R_{IJ}}x_{IJ}^{\alpha}(\delta_{IK}-\delta_{JK})\label{eqn:fk}
\end{eqnarray}
and
\begin{eqnarray}
\frac{\p \sa_{IJ}}{\p  R_{K}^{\alpha}} &=& \bigg(\frac{\p \sa_{I}^{(0)}}
{\p  \rho_{I}^{(0)}}
+\frac{\p \sa_{IJ}^{(1)}}{\p  \rho_{I}^{(0)}}\bigg)
 \frac{\p \rho_{I}^{(0)}}
{\p  R_{K}^{\alpha}} + \frac{\p \sa^{(1)}_{IJ}}{\p\rho^{(1)}_{IJ}} \frac{\p \rho_{IJ}^{(1)}}
{\p  R_{K}^{\alpha}} \\
\frac{\p \rho_{I}^{(0)}}
{\p  R_{K}^{\alpha}} &=& \sum_{L(I)}\frac{\p f_{IL}}{\p R_{IL}}x^{\alpha}_{IL}(\delta_{IK}-\delta_{LK}) \\
\frac{\p \rho_{IJ}^{(1)}}{\p  R_{K}^{\alpha}} &=& \sum_{L(I)}\frac{1}{R_{LI}}(\dd_{LK}-\dd_{IK})(\dd^{\alpha\beta}
-x_{LI}^{\beta}x_{LI}^{\alpha})x_{JI}^{\beta}f_{LI}A_{IJL}\nonumber\\
&+& \sum_{L(I)}\frac{1}{R_{JI}}(\dd_{JK}-\dd_{IK})(\dd^{\alpha\beta}-x_{JI}^{\beta}x_{JI}^{\alpha})
x_{LI}^{\beta}f_{LI}A_{IJL}\nonumber\\
&+& \sum_{L(I)}x_{LI}^{\beta}x_{JI}^{\beta}\frac{\p f_{LI}}{R_{LI}}x_{LI}^{\alpha}(\dd_{LK}-\dd_{IK})A_{IJL}
\end{eqnarray}
where $A_{IJK} = \Theta({\mathbf x}_{IJ}\cdot{\mathbf x}_{IK})$.
The notation $ \sum_{L(I)}$ has been introduced to indicate that the summation is over all ions
$L$ that are neighbours of $I$. This is necessary for practical implementation
due to the truncation of interactions and to avoid summing over all the particles.

Expanding equation~\ref{eqn:fk} we get
\begin{eqnarray}
F_{K}^{\alpha} &=&
% **1**
-\sum_{J(K)} U^{\text{\tiny SR}}_{KJ}\frac{\p g_{KJ}}{R_{KJ}}x_{KJ}^{\alpha} \nonumber\\
% **2**
&-& \sum_{J(K)} \frac{\p U^{\text{\tiny SR}}_{KJ}}{\p L_{KJ}}g_{KJ}x_{KJ}^{\alpha} \nonumber\\
% **3**
&+& \sum_{I(K)} \frac{\p f_{KI}}{\p R_{KI}} x_{KI}^{\alpha}\sum_{J(I)}  \frac{\p U^{\text{\tiny SR}}_{IJ}}{\p L_{IJ}}g_{IJ}
\bigg(\frac{\sa^{(0)}_{I}}{\p \rho_{I}^{(0)}} + \frac{\sa_{IJ}^{(1)}}{\p \rho_{I}^{(0)}}\bigg) \nonumber\\
% **4**
&+& \Bigg(\sum_{J(K)} \frac{\p U^{\text{\tiny SR}}_{KJ}}{\p L_{KJ}}g_{KJ}\bigg(\frac{\sa^{(0)}_{K}}{\p \rho_{K}^{(0)}}
+ \frac{\sa_{KJ}^{(1)}}{\p \rho_{K}^{(0)}}\bigg)\Bigg)\Bigg(\sum_{I(K)}
\frac{\p f_{KI}}{\p R_{KI}} x_{KI}^{\alpha}\Bigg) \nonumber\\
% **5**
&-& \sum_{I(K)}\frac{f_{KI}}{R_{KI}}\sum_{J(I)^{(1)}}
\frac{\p U^{\text{\tiny SR}}_{IJ}}{\p L_{IJ}}g_{IJ}
\frac{\p \sa_{IJ}^{(1)}}{\p \rho_{IJ}^{(1)}}x_{IJ}^{\alpha}A_{IJK}  \nonumber\\
% **6**
&+& \sum_{I(K)}\frac{x_{KI}^{\alpha}}{R_{KI}}f_{KI}
\sum_{J(I)} \frac{\p U^{\text{\tiny SR}}_{IJ}}{\p L_{IJ}}g_{IJ}
\frac{\p \sa_{IJ}^{(1)}}{\p \rho_{IJ}^{(1)}}x_{KI}^{\beta}x_{IJ}^{\beta}A_{IJK} \nonumber\\
% **7**
&+& \sum_{I(K)}\frac{\p U^{\text{\tiny SR}}_{KI}}{\p L_{KI}}g_{KI}\frac{\p \sa_{KI}^{(1)}}{\p \rho_{KI}^{(1)}}x_{KI}^{\alpha}
\sum_{J(K)}\frac{f_{JK}}{R_{JK}}A_{KIJ} \nonumber\\
% **8**
&-& \sum_{I(K)}\frac{\p U^{\text{\tiny SR}}_{KI}}{\p L_{KI}}g_{KI}\frac{\p \sa_{KI}^{(1)}}{\p \rho_{KI}^{(1)}}
\sum_{J(K)}\frac{f_{JK}}{R_{JK}}x_{KJ}^{\beta}x_{KI}^{\beta}x_{KJ}^{\alpha}A_{KIJ}\nonumber\\
% **9**
&-&\sum_{I(K)} \frac{\p U^{\text{\tiny SR}}_{KI}}{\p L_{KI}}
\frac{g_{KI}}{R_{KI}}
\frac{\p \sa_{IK}^{(1)}}{\p \rho_{IK}^{(1)}}
\sum_{J(I)}x_{IJ}^{\alpha}f_{IJ}A_{IKJ} \nonumber\\
% **10**
&+& \sum_{I(K)}\frac{\p U^{\text{\tiny SR}}_{KI}}{\p L_{KI}}g_{KI}\frac{\p\sa_{IK}^{(1)}}{\p \rho_{IK}^{(1)}}
\frac{x_{KI}^{\alpha}}{R_{KI}}
\sum_{J(I)}x_{KI}^{\beta}x_{IJ}^{\beta}f_{IJ}A_{IKJ} \nonumber\\
% **11**
&+& \sum_{I(K)}\frac{\p U^{\text{\tiny SR}}_{KI}}{\p L_{KI}}\frac{g_{KI}}{R_{KI}}\frac{\p \sa_{KI}^{(1)}}{\p \rho_{KI}^{(1)}}
\sum_{J(K)} x_{KJ}^{\alpha}f_{KJ}A_{KIJ} \nonumber\\
% **12**
&-&\sum_{I(K)}\frac{\p U^{\text{\tiny SR}}_{KI}}{\p L_{KI}}\frac{g_{KI}}{R_{KI}}\frac{\p \sa_{KI}^{(1)}}{\p \rho_{KI}^{(1)}}
x_{KI}^{\alpha}
\sum_{J(K)}x_{KI}^{\beta}x_{KJ}^{\beta}f_{KJ}A_{KIJ} \nonumber\\
% **13**
&-&\sum_{I(K)}\frac{\p f_{KI}}{\p R_{KI}}x_{KI}^{\alpha}\sum_{J(I)}\frac{\p \sa_{IJ}^{(1)}}{\p \rho_{IJ}^{(1)}}
 \frac{\p U^{\text{\tiny SR}}_{IJ}}{\p L_{IJ}}g_{IJ}x_{KI}^{\beta}x_{IJ}^{\beta}A_{IJK} \nonumber\\
% **14**
&+&\sum_{I(K)}\frac{\p U^{\text{\tiny SR}}_{KI}}{\p L_{KI}}g_{KI}\frac{\p \sa_{KI}^{(1)}}{\p \rho_{KI}^{(1)}}
\sum_{J(K)}x_{KJ}^{\beta}x_{KI}^{\beta}x_{KJ}^{\alpha}\frac{\p f_{KJ}}{\p R_{KJ}}A_{KIJ}
\end{eqnarray}
In the derivation we have made the further assumptions that $f_{IJ}=f_{JI}$,
$U^{\text{\tiny SR}}_{IJ}=U^{\text{\tiny SR}}_{JI}$ and $g_{IJ}=g_{JI}$.
\clearpage
\end{section}

% LIQUID TABLE

\clearpage
\newpage

\section*{TABLES}
\begin{table}[tbh]
 \centering
 \caption{The fit to the {\em ab initio} data for the different potential forms}
\begin{tabular}{|c||c|c|c||c|c|c||c|c|c||}
\hline
& \multicolumn{3}{|c|}{$300$K crystal}
& \multicolumn{3}{|c|}{$2000$K crystal}
& \multicolumn{3}{|c|}{$3000$K Liquid}  \\
\cline{2-10}
&$\Delta F$&$\Delta S$&$\Delta E$&$\Delta F$&$\Delta S$&$\Delta E$&$\Delta F$&$\Delta S$&$\Delta E$\\
\hline
 A &9.3&5.0&25.5&13.7 &3.3 &15.5&25.1&4.8&52.4\\
 B &6.9&5.2&23.8&9.0&6.2&17.8&17.5&5.6&23.6\\
 C &10.4&39.1 &5.9&13.6&51.7&164.8&32.2&58.2&69.2\\
 D & 3.4& 0.6& 3.0&6.8 &0.3 &9.8&17.1&0.3&10.5\\
 E &12.8&0.1 &59.0&10.2&0.1 &18.9&9.6&0.0&17.7\\
\hline
    \end{tabular}
\label{table:pot_test}
 \end{table}
%%%%%%%%%%%%%%%%%%%%%%%%%%%%%%%%%%%%%
 \begin{table}[tbh]
  \centering
  \caption{The fit to the LDA {\em ab initio} data for the liquid (F) and solid (G) potentials used in the
 calculation of the melting slope.}
 \begin{tabular}{|c||c|c|c||c|c|c||}
 \hline
 & \multicolumn{3}{|c|}{$3000$K Crystal}
 & \multicolumn{3}{|c|}{$3000$K Liquid} \\
 \cline{2-7}
 &$\Delta F$&$\Delta S$&$\Delta E$&$\Delta F$&$\Delta S$&$\Delta E$\\
 \hline
  F &9.6&0.1&10.8&10.4&0.2&10.2\\
  G &6.2&0.3&10.6&44.0&2.3&54.0\\
 \hline
     \end{tabular}
 \label{table:pot_test2}
 \end{table}
%%%%%%%%%%%%%%%%%%%%%%%%%%%%%%%%%%%%%
\begin{table}
\caption{Parameters for potential F (atomic units)}
 \centering
\begin{tabular}{|c||c|c|c||}
\hline
Parameter &Mg-Mg&Mg-O&O-O\\
\hline
\hline
 $b_{\text{pol}}$\footnotemark[1] & - & $1.55713$ & $4.01338$ \\
 $c_{\text{pol}}$\footnotemark[1] & - & $-1.28035$ & $31.93748$ \\
 $C^{(1)} $\footnotemark[2] & - & $7.63066\times 10^{1}$ & $1.58612\times 10^{3}$ \\
 $C^{(2)} $\footnotemark[2] & - & $2.01709$ & $2.42329$ \\
 $C^{(4)} $\footnotemark[3] & - & $-4.09954\times 10^{-2}$ & $2.248341\times 10^{-2}$ \\
 $A$\footnotemark[4] & $4.10638\times 10^{12}$ & $  1.82404\times 10^{2}$ & $-2.78524\times 10^{4}$ \\
 $\alpha$\footnotemark[4] & $ 7.29702$& $ 2.22211$& $2.98764$ \\
 $B$\footnotemark[4] & $-3.48800\times 10^{13}$ & $-2.00799\times 10^{3}$ & $4.00143\times 10^{3}$ \\
 $\beta$\footnotemark[4] & $7.80852$ & $3.71852$ & $ 2.44883$ \\
\hline
    \end{tabular}
\label{table:parameters_liquid}

 \centering
 \centering
\begin{tabular}{|c||c|c||}
\hline
Parameter &Mg&O\\
\hline
\hline
 $q$ & $1.44831$ & $-1.44831$ \\
 $\alpha$ & - & $14.25305$ \\
 $C_{(3)}$\footnotemark[5] & - & $-1.20698\times 10^{-2}$ \\
 $\epsilon_{1}$\footnotemark[6] & - & $-1.44105\times 10^{7}$ \\
 $\epsilon_{2}$\footnotemark[6] & - & $ -5.14776$ \\
 $\epsilon_{3}$\footnotemark[6] & - & $1.11436\times 10^{8}$ \\
 $\epsilon_{4}$\footnotemark[6] & - & $ 7.20899$ \\
\hline
    \end{tabular}
\label{table:parameters}

\begin{tabular}{|c||c|c|c||}
\hline
Parameter &Mg-Mg&Mg-O&O-O\\
\hline
\hline
 $C^{6}$\footnotemark[7] & - & $-1.01505\times 10^{3}$  & $3.24685\times 10^{4}$ \\
 $C^{8}$\footnotemark[7]& - & $9.16150\times 10^{3}$ , $-6.70301\times 10^{6}$ & $-5.93805 \times 10^{6}$ \\
 $b_{\text{tt}}^{6}$\footnotemark[7] & - & $2.51113\times 10^{-1}$  & $-1.60446 \times 10^{-1}$ \\
 $b_{\text{tt}}^{8}$\footnotemark[7] & - & $3.78280 \times 10^{-1}$ , $9.07552\times 10^{-2}$ & $-7.45819\times 10^{-2}$ \\
\hline
\label{table:parameters_disp}
    \end{tabular}
\footnotetext[1]{Defined as in equation 4 of reference~\onlinecite{us_silica}}
\footnotetext[2]{See equation~\ref{eqn:fij2}}
\footnotetext[3]{See equation~\ref{eqn:sigmaij}}
\footnotetext[4]{See equation~\ref{eqn:ab_def}}
\footnotetext[5]{See equation~\ref{eqn:sigmai}}
\footnotetext[6]{See equation~\ref{eqn:selfenergy2}}
\footnotetext[7]{Defined as in Ref.~\onlinecite{madden_compress}.}

\end{table}
\newpage

\begin{table}
\caption{Parameters for potential G (atomic units)}
 \centering
\begin{tabular}{|c||c|c|c||}
\hline
Parameter &Mg-Mg&Mg-O&O-O\\
\hline
\hline
 $b_{\text{pol}}$ & - & $1.65744$ & $4.01338$ \\
 $c_{\text{pol}}$ & - & $-1.35136$ & $31.93748$ \\
 $C^{(1)} $ & - & $1.00281$ & $1.72179\times 10^{-5}$ \\
 $C^{(2)} $ & - & $0.65944$ & $0.19425$ \\
 $C^{(4)} $ & - & $-3.18047\times 10^{-2}$ & $-2.93661\times 10^{-2}$ \\
 $A$ & $1.39895\times 10^{11}$ & $ 2.16223\times 10^{2}$ & $-5.03851\times 10^{4}$ \\
 $\alpha$ & $6.72904$ & $2.23873$ & $3.09196$ \\
 $B$ & $-2.50649\times 10^{13}$ & $-4.21223\times 10^{2}$ & $4.29645\times 10^{3}$ \\
 $\beta$ & $7.98372$ & $ 3.16438$ & $ 2.43995$ \\
\hline
    \end{tabular}
\label{table:parameters_solid}

 \centering
\begin{tabular}{|c||c|c||}
\hline
Parameter &Mg&O\\
\hline
\hline
 $q$ & $1.48077$ & $-1.48077$ \\
 $\alpha$ & - & $10.40993$ \\
 $C_{(3)}$ & - & $-1.33660\times 10^{-2}$ \\
 $\epsilon_{1}$ & - & $-5.59871\times 10^{8}$ \\
 $\epsilon_{2}$ & - & $ 7.61483\times 10^{-1}$ \\
 $\epsilon_{3}$ & - & $4.89669\times 10^{8}$ \\
 $\epsilon_{4}$ & - & $4.59914\times 10^{-1}$ \\
\hline
    \end{tabular}
\label{table:parameters_solid2}
\end{table}

\clearpage
\newpage

\section*{Figures}
\begin{figure}[!p]
\begin{center}
\includegraphics[width=9cm,height=6.75cm]{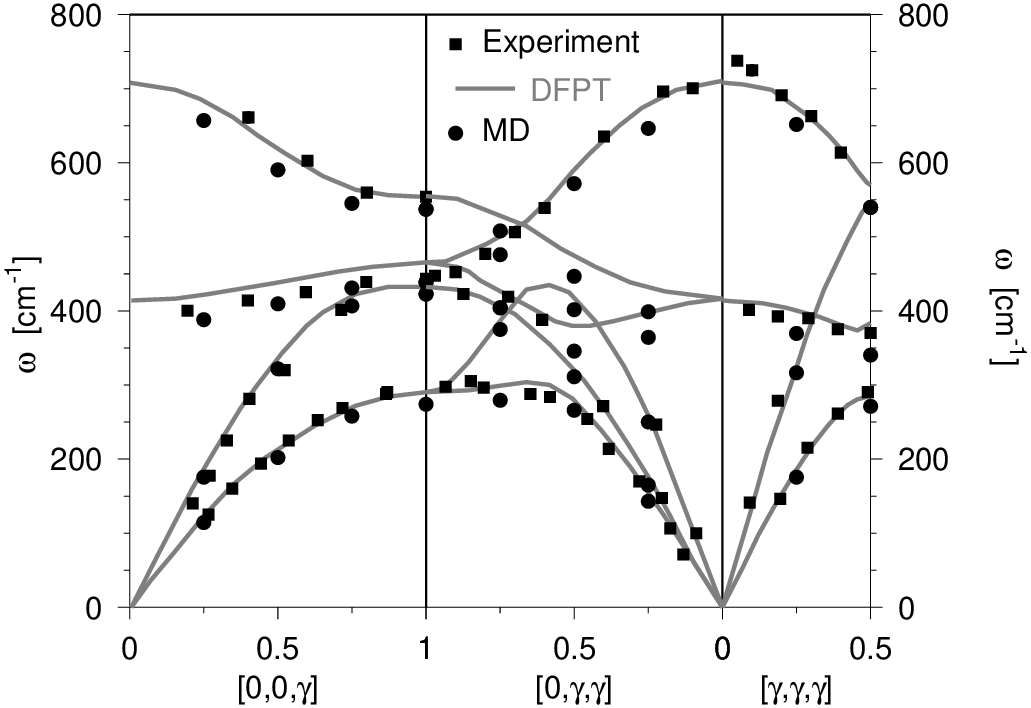}
\end{center}
\caption{}
\label{fig:phonon_full}
\end{figure}

\begin{figure}[!p]
\begin{center}
\includegraphics[width=9cm,height=6.75cm]{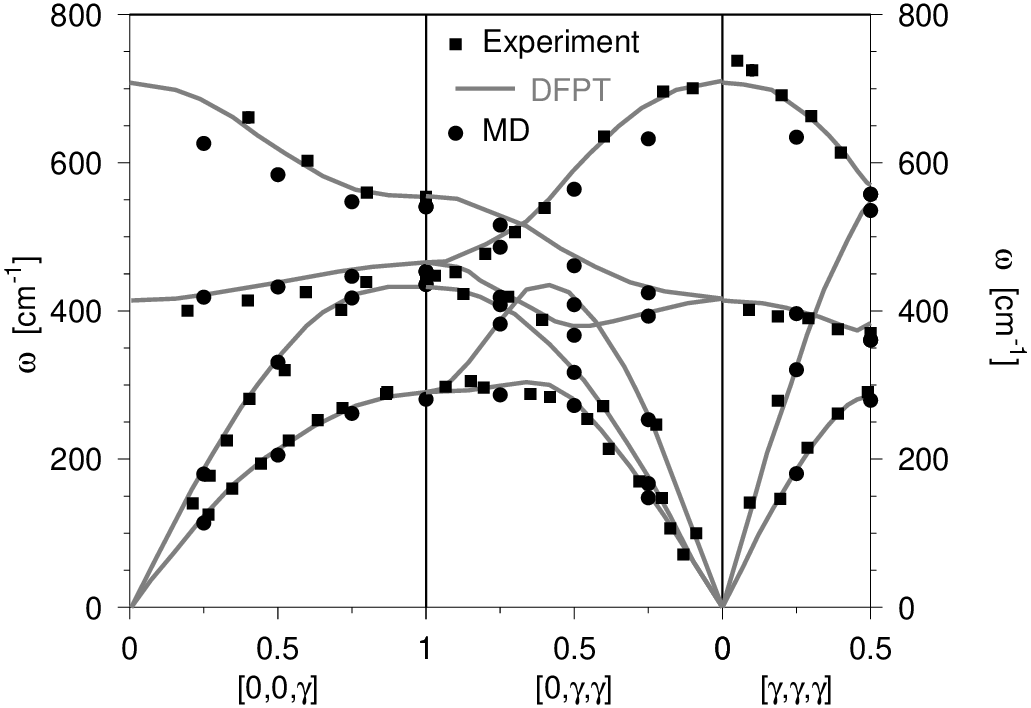}
\end{center}
\caption{}
\label{fig:phononF}
\end{figure}

\begin{figure}[!p]
\begin{center}
\includegraphics[width=9cm,height=9cm]{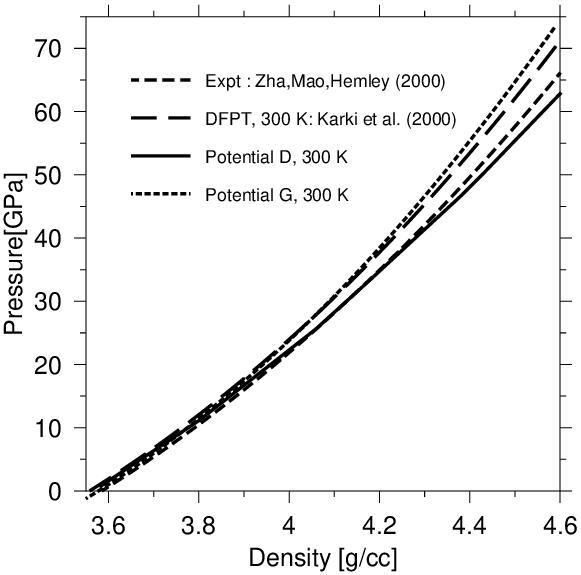}
\caption{}
\label{fig:eos}
\end{center}
\end{figure}

\begin{figure}[!p]
\begin{center}
\includegraphics[width=9cm,height=9cm]{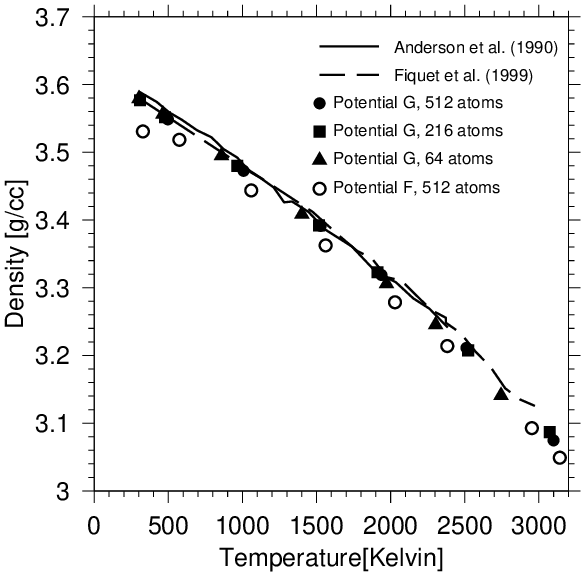}
\caption{}
\label{fig:thermal}
\end{center}
\end{figure}

\begin{figure}[!p]
\begin{center}
\includegraphics[width=9cm,height=9cm]{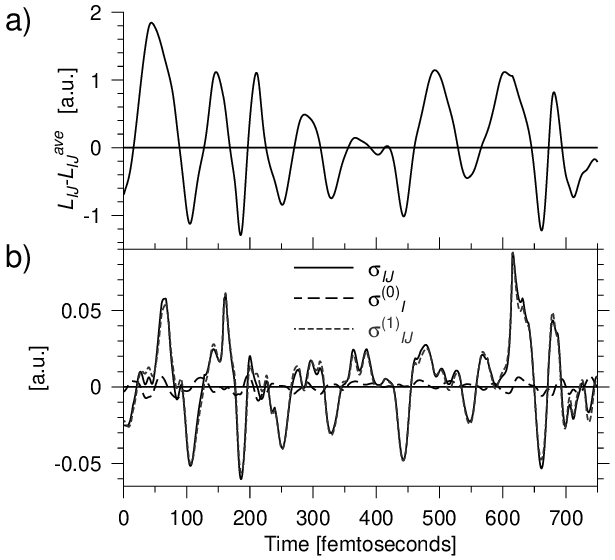}
\caption{}
\label{fig:radius}
\end{center}
\end{figure}

\begin{figure}[!p]
\begin{center}
\includegraphics[width=9cm,height=4.5cm]{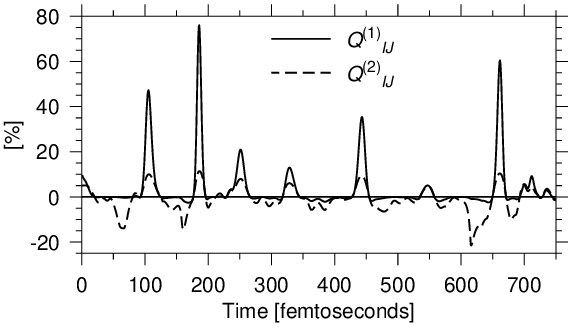}
\end{center}
\caption{}
\label{fig:force}
\end{figure}

\begin{description}
\item{Fig. \ref{fig:phonon_full}}
The phonon dispersions of MgO as calculated with the full polarizable and distortable-ion model parametrized in
the crystal under ambient conditions (Potential D)
compared with experiment\cite{sangster} and with the density functional perturbation theory results of
Karki {\em et al.}\cite{karki}

\item{Fig. \ref{fig:phononF}}
The phonon dispersions of MgO as calculated with the full polarizable and distortable-ion model parametrized in
the crystal at $3000$K (Potential G)
compared with experiment\cite{sangster} and with the density functional perturbation theory results of
Karki {\em et al.}\cite{karki}
\item{Fig. \ref{fig:eos}}
The pressure of MgO as a function of density (at $300$ K) compared to experiment\cite{mao} and
density functional perturbation theory calculations \cite{karki}.
MD simulations used the full model potential parametrized at ambient conditions and simulation cells containing
$512$ atoms.
\item{Fig. \ref{fig:thermal}}
The density of MgO as a function of
temperature (at zero pressure) compared to experiment\cite{thermal1,thermal2} for potentials
G and F. Three different simulation cell sizes are used to check for finite size effects.
\item{Fig. \ref{fig:radius}}
a) $L_{IJ}-\overline{L_{IJ}}$ as a function of time, where $\overline{L_{IJ}}\approx4.66$ a.u. is the
average over the trajectory of $L_{IJ}$, the inter-{\em membrane} distance (see section~\ref{section:our_aim}).
b) $\sigma_{IJ}-\overline{\sigma_{IJ}}$,$\sigma^{(1)}_{IJ}-\overline{\sigma^{(1)}_{IJ}}$, and
$\sigma^{(0)}_{I}-\overline{\sigma^{(0)}_{I}}$. $I$ and $J$ are neighbouring anion and cation respectively.
\item{Fig. \ref{fig:force}}
$Q^{(1)}_{IJ}$ and $Q^{(2)}_{IJ}$ 
(see equations~\ref{eqn:q1} and~\ref{eqn:q2})as a function of time along the same trajectory 
shown in figure~\ref{fig:radius}
\end{description}

%%%%%%%%%%%%%%%%%%%%%%%%%%%%%%%%%

\newpage

\newpage

\newpage

\newpage

\newpage


\begin{references}
\bibitem{sample}
Averages are taken over a large number of sample configurations
that have not been used in the optimization of parameters.
\bibitem{eos}
We use a much higher plane wave cutoff and this may become increasingly relevant at 
high pressures. We also sample the  Brillouin zone in a different way.
\end{references}
\end{document}